\documentclass{PoS}

\usepackage{heppennames2}

\usepackage{lineno}

\newcommand{\pel}{\ensuremath{\mathcal{P}_{\Pem}}}
\newcommand{\ppol}{\ensuremath{\mathcal{P}_{\Pep}}}
\newcommand{\Leff}{\ensuremath{\mathcal{L}_{\text{eff}}}}
\newcommand{\Peff}{\ensuremath{\mathcal{P}_{\text{eff}}}}
\newcommand{\Lum}{\ensuremath{\mathcal{L}}}
\newcommand{\LR}{\ensuremath{\text{LR}}}

\newcommand{\ALR}{\ensuremath{A_\LR}}

\graphicspath{{./plots/}}

\title{The Role of Positron Polarization for the initial 250 GeV stage
of the ILC}

\ShortTitle{Role of Positron Polarization for initial stage ILC-250}

\author{
  \speaker{J\"urgen Reuter} ~~~ \\
  on behalf of the Physics Working Group of the LCC Collaboration \\
  DESY, Hamburg, Germany\\
  E-mail: \email{juergen.reuter@desy.de}
}

\abstract{
The International Linear Collider is now proposed with a staged
machine design, with the first stage at $\sqrt{s}=250$ GeV and an
integrated luminosity goal of 2 ab${}^{-1}$. One of the questions for
the machine design is the importance of positron polarization. In this
report, we review the impact of positron polarization on the physics
goals of the 250 GeV stage of the ILC and demonstrate that positron
polarization has distinct advantages. 
}

\FullConference{39th International Conference on High Energy Physics\\
		4-11 July 2018\\
		Seoul, Korea}

\begin{document}

\section{Introduction}

The International Linear Collider (ILC) is a linear $\Pep\Pem$
collider~\cite{ILC_TDR} 
with a staged operation plan: its initial stage has a center-of-mass
energy of 250 GeV (20 km length)~\cite{Barklow:2015tja} and will
deliver an integrated luminosity of up to 250 fb${}^{-1}$/year. Later
upgrade options include energies of 350 GeV, 500 GeV (31 km length) or
even 1 TeV (50 km length). The ILC is based on superconducting RF
cavities with 31.5 MV/m design gradient. In the baseline design,
beams with 80 \% electron polarization and 30 \% positron polarization
are foreseen. The project is at the moment under consideration by the
Japanese government with a proposed cite in Kitakami in the Iwate
prefecture in Northern Japan. The ILC as an electron-positron-collider
offers an incredible physics program due to the very clean
experimental setup with a well-defined initial state, pure electroweak
production and a triggerless operation. This program comprises
per-cent and even per-mil-level measurements in the Higgs and top
sectors with the possibilities for discoveries of physics beyond the
Standard model. The ILC offers the only option to search for light
electroweak states that are elusive at hadron colliders. The physics
case for the ILC has been summarized by the LCC Physics Working Group
in~\cite{Fujii:2015jha,Fujii:2017ekh} and has been updated for the
staged scenario in~\cite{Fujii:2017vwa}. A polarized electron beam
with up to 80\% polarization is part of the baseline design in every
case. In the latest staging report~\cite{Evans:2017rvt}, there are two
different technological options for the positron source: (1) according
to the TDR baseline design, a polarized source using a super-conducting
helical undulator of 231 m length for a 125 GeV electron beam
producing a highly collimated polarized photon beam shot on a rotating
target, which creates polarized positrons. A spin-rotating solenoid
allows any rotation of the polarization vector, in particular both
longitudinal and also transversal polarization modes. This option
needs the full 125 GeV beam for commissioning. (2) the alternative,
more "conventional" design uses a normal-conducting linear
accelerator. This option produces unpolarized positrons only. It does
not need the full electron beam energy for commissioning, however,
leads to a more complicated bunch structure, which possibly needs a
second positron damping ring. The technology choice will most likely
be made only after an approval of the machine. In this contribution,
we discuss why $\Pep$ polarization is such an important asset of the
ILC physics program. 

\section{The Surplus of Positron Polarization}
\label{sec:posipol}

The two main rationales of polarization are higher rates for signal
samples, which saves on running time and operation costs, and it
enhances signal-to-background ratios by simultaneously suppressing
backgrounds. We will first give a general argument for the enhancement
of the physics potential with $\Pep$ polarization, and will then
discuss more specific examples.

Beam polarization fractions ($\pel$, $\ppol$) are defined as the
difference of right and left-handed positron numbers normalized to the
sum. First, cross section rates are proportional to the effective
luminosity, given in terms of the beam luminosity as $\Leff := \frac12
\left( 1 - \pel\ppol \right) \Lum$. For the default ILC setup, $\pel =
-0.8$, $\ppol = +0.3$, this is $0.62\,\Lum$, while it drops to
$0.5\,\Lum$ without $\Pep$ polarization, which amounts to 24\% loss
in luminosity. Second, for the most powerful model-independent searches
for physics beyond the SM, by parameterizing deviations from the SM
in terms of dimension-6 operators in an Effective Field Theory (EFT)
expansion, the resolution power for its operator coefficients is
greatly enhanced by looking for left-right asymmetries, $\ALR =
\left(\sigma_\LR - \sigma_{\text{RL}}\right) / \left(\sigma_\LR +
\sigma_{\text{RL}} \right)$. The crucial quantity in these
measurements is the effective polarization, $\Peff = ( \pel -
\ppol ) / ( 1 - \pel\ppol )$, whose drop from $-0.89$
for the standard ILC setup to $-0.8$ without $\Pep$ polarization
translates to a 10\% loss of analyzing power for the
$A_{\text{LR}}$. Besides the increase of $\Peff$ with $\Pep$
polarization, the latter also reduces the sensitivity of $\Peff$ to
beam polarization uncertainties. The role of $\Pep$ polarization in
$\Pem\Pep$ collisions is reviewed in~\cite{MoortgatPick:2005cw}. 

Important physics examples degraded without $\Pep$
polarization are (1) $s$-channel processes ($\gamma/Z$ exchange) where
only $\sigma_{\text{LR}}$ and $\sigma_{\text{RL}} \neq 0$. Hence,
using the unpolarized cross section, $\sigma_0$, the polarized cross
section is $\sigma(\pel,\ppol) = 2 \sigma_0 \Leff/\Lum \left[ 1 -
\Peff\ALR \right]$. The Higgsstrahlung process and hence the whole 250
GeV Higgs program as well as di-fermion processes like $\Pem\Pep \to
\PQb\PAQb$ are in this category; then, (2) $t$-channel $W/\nu$
exchange for which only $\sigma_\LR \neq 0$, and $\sigma(\pel,\ppol) =
2 \sigma_0 \Leff/\Lum \left[ 1 - \Peff\right]$. With $\Pep$
polarization, these processes, $W^+W^-$ production and vector boson
fusion into Higgs, have 30\% larger cross section and also 30\% less
background. Likewise, a process like single-$W$ production
(which allows a very precise determination of the $W$ mass) has
only one non-vanishing polarization in either the initial electron or
positron (depending on the sign of the $W$), and profits similarly
from $\Pep$ polarization. In direct BSM searches, all four
polarization combinations can contribute, such that there are more
varieties, discussed below.

For a detailed quantization of the surplus of $\Pep$ polarization,
Ref.~\cite{Fujii:2017vwa} assumed two scenarios for 2 ab${}^{-1}$ at 250
GeV, "$\Pep$ pol." with $|\pel| = 80 \%$,
$|\ppol| = 30\%$ and a running scenario with 45\% in the
$(\Pem,\Pep)$ polarization configuration $(+,-)$, 45\% in $(-,+)$, and
5\% in each of $(+,+)$ and $(-,-)$, respectively. This is compared to
the scenario "no $\Pep$ pol." with $|\pel| = 80 \%$,
$|\ppol| = 0\%$ and a running scenario with 50\% each in
$(+,\cdot)$ and in $(-,\cdot)$, respectively ($\cdot\equiv$
unpol.). In general, a combination of data samples with different
polarizations gives a higher sensitivity than a single sample, it
allows to study like-sign polarization sets, and samples with
different $\Pep$ polarization allow for significant reduction of
systematic uncertainties by using four different data sets (even nine
different sets including transversal polarization).

We now give three examples to show the relevance of $\Pep$
polarization in SM precision measurements: (1) for a minimization of
systematic uncertainties of total cross section and left-right
asymmetry measurements, $\Pep$ polarization is crucial. While
electron polarization can always be determined to the sub-permil
level, without $\Pep$ polarization systematic uncertainties in Higgs
measurements are up to a factor of five larger without $\Pep$
polarization~\cite{robert_thesis}. Even when using positron
polarimeters, the necessity to experimentally confirm vanishing
$\Pep$ polarization leads to a factor of two to three larger
uncertainties for single-$W$, $WW$ measurements and two-fermion
measurements. An independent consistency check on systematics is
only possible with $\Pep$ polarizations. (2) For the search for
deviations from the SM in terms of an EFT, in $\Pem\Pep \to W^+W^-$,
the most general EFT in that measurement has 14 complex parameters,
which can only be extracted using $\Pep$ polarization. Also, the EFT
Wilson coefficient determination of trilinear couplings in $\Pem\Pep
\to Z\gamma$ needs $\Pep$ polarization, so measurements of
of general SMEFT coefficients are only possible with $\Pep$
polarization. (3) As a third example, we discuss the importance of
$\Pep$ polarization in the Higgs precision program of the ILC, where
it enhances the $ZH$ cross section from 420,000 to 500,000 Higgs
bosons with 2 ab${}^{-1}$ at 250 GeV, which is a reduction in running
costs by roughly 19\%. For the EFT Higgs coupling fit, having no
$\Pep$ polarization amounts to a degradation of up to 6 \%, mainly
from statistics~\cite{Barklow:2017suo}. For the systematics, the
luminosity uncertainty depends much on the $\Pep$ polarization, again,
for the polarization uncertainty, there is a bias from the
polarimeter, and different polarization samples allow an {\em in-situ}
reduction of background uncertainty by a determination of the
background from the polarization combinations, which disfavors the
signal. In addition, $\Pep$ polarization allows new tests of the
whole EFT framework by overconstraining the setup.

Finally, we discuss the impact of $\Pep$ polarization on BSM
searches. There is a general paradigm, that a polarized positron
source could be built after a possible new physics discovery for
discrimination, however, such a separation is not necessarily
possible. Most likely, a large data set is needed to establish
a 3$\sigma$ evidence or 5$\sigma$ discovery, and $\Pep$ polarization
would very early on provide a handle on the significance. Discoveries
could even be missed without $\Pep$ polarization: examples are
contact interactions ($Z^\prime$), where $\Pep$ polarization enhances
the reach for new physics scales by 30\% by using four different
polarization samples, the search for heavy leptons in $\Pem\Pep \to
W^+W^-$ needs double polarization asymmetries, which is only possible
with $\Pep$ polarization, and the search for light
pseudo-Nambu-Goldstone  bosons in $\Pem\Pep \to \PQb\PAQb\phi$ needs
$\Pep$ polarization for a determination of the quantum
numbers~\cite{Kilian:2004pp}, and in the search for invisible
particles the possibility of using LR/RL versus LL/RR samples allows
to separate signal from backgrounds, and in EFT dark matter searches
$\Pep$ polarization saves up to 2.5 years of run time.

\section{Conclusions}

The ILC in its 250 GeV stage offers an indispensable physics program
with its high-precision Higgs program, model-independent searches of
electroweak states and direct dark matter searches. Positron
polarization crucially enhances the physics potential by an
improvement of signal-to-background ratios, a safeguard against
systematic uncertainties and by allowing much higher sensitivity
through a combination of different polarization data samples. Several
of these points lead to savings on running costs of up to 20 \%. These
issues should be carefully considered when the technology choice for
the positron source is made.  


\end{document}